\documentclass[sigconf]{acmart}

\usepackage{array}
\usepackage{cleveref}
\usepackage{tikz}
\usetikzlibrary{shapes,arrows, matrix, decorations.pathreplacing, arrows.meta, positioning}

\copyrightyear{2026}
\acmYear{2026}
\setcopyright{othergov}
\acmConference[PASC '26]{Platform for Advanced Scientific Computing Conference}{June 29-July 01, 2026}{Bern, Switzerland}
\acmBooktitle{Platform for Advanced Scientific Computing Conference (PASC '26), June 29-July 01, 2026, Bern, Switzerland}
\acmDOI{10.1145/3815572.3815754}
\acmISBN{979-8-4007-2734-4/2026/06}

\begin{document}

\title{A Comparison of Massively Parallel Performance Portable Particle-in-Cell schemes for electrostatic kinetic plasma simulations}

\author{Sonali Mayani}
\email{sonali.mayani@psi.ch}
\orcid{0000-0002-3294-4650}
\affiliation{
 \institution{PSI Center for Scientific Computing, Theory and Data}
 \city{Villigen PSI}
 \country{Switzerland}}
\affiliation{
  \institution{ETH Zürich}
  \city{Zürich}
  \country{Switzerland}}
  
\author{Paul Fischill}
\email{pfischill@ethz.ch}
\affiliation{
 \institution{ETH Zürich}
 \city{Zürich}
 \country{Switzerland}}

\author{Sriramkrishnan Muralikrishnan}
\email{s.muralikrishnan@fz-juelich.de}
\affiliation{
  \institution{Jülich Supercomputing Centre, Forschungszentrum Jülich GmbH}
  \city{Jülich}
  \country{Germany}}

\author{Andreas Adelmann}
\email{andreas.adelmann@psi.ch}
\affiliation{
 \institution{PSI Center for Scientific Computing, Theory and Data}
 \city{Villigen PSI}
 \country{Switzerland}}

\renewcommand{\shortauthors}{Mayani et al.}

\begin{abstract}

We compare different Poisson solvers within the context of an electrostatic Vlasov-Poisson system. 
These schemes are implemented as part of the IPPL (Independent Parallel Particle Layer) library \cite{matthias_frey_2024_10878166}, which provides performance portable and dimension independent building blocks for scientific simulations requiring particle-mesh methods, with Eulerian (mesh-based) and Lagrangian (particle-based) approaches.

The simulation used to compare the performance and portability of the schemes is Landau damping, part of a set of mini-applications implemented to benchmark and showcase the capabilities of the IPPL library \cite{muralikrishnanScalingPerformancePortability2024}. We use grid-sizes of $512^3$ and $1024^3$ with 8 particles per cell, running with different algorithms in the solve phase of the Particle-in-Cell (PIC) loop: a Fast Fourier Transform (FFT) pseudo-spectral solver, a matrix-free finite difference Preconditioned Conjugate Gradient (PCG) solver, and a matrix-free Finite Element (FEM) solver. We also compare these PIC schemes to the novel Particle-in-Fourier (PIF) scheme, which performs interpolations using non-uniform FFTs thereby avoiding a grid in the real space. We obtain results on different computing architectures, such as AMD GPUs (LUMI at CSC), and Nvidia GPUs (Alps at CSCS and JUWELS Booster at Jülich Supercomputing Center), showcasing portability. In terms of absolute time the FFT solver is advantageous, but is limited in its applicability. All other field solvers in the PIC scheme are an order-of-magnitude more expensive in terms of time, but scale similarly to the FFT case in the electrostatic PIC context. The PIF scheme serves as a high fidelity alternative to standard PIC, and while it is costlier than the FFT-based PIC scheme, it shows excellent scalability on all the architectures. 

\end{abstract}



\maketitle

\section{Introduction}

The field of High-Performance Computing (HPC) is delving into heterogeneous architectures to increase performance. As power constraints and Moore's law start to limit the scales one can reach with traditional CPU architectures, the use of accelerators such as GPUs has become mandatory. New exascale machines for scientific computing increasingly combine heterogeneous CPUs and GPUs from multiple vendors. Looking at only the top 5 supercomputers of the TOP500 list as of November 2025, one can already find a multitude of architectures: AMD CPUs and GPUs, Intel CPUs and GPUs, and Nvidia GPUs.

To take full advantage of these machines, or even be able to run on such heterogeneous architectures, it has become imperative for computational simulations to be massively parallelized and portable. Simulation codes need to be hardware independent so that they can run on novel super-computing clusters without needing to write separate codes for each architecture. Scientific programming now needs to be quickly adaptable to these new architectures in order to accelerate applications. To fully exploit the extremely high floating-point throughput of modern supercomputers, simulation codes must exhibit massive parallelism and high arithmetic intensity. This is essential to enable simulations at significantly higher spatial and temporal resolutions, which in turn allows finer-scale physical phenomena to be resolved.

In this work we are interested in solving the electrostatic Vlasov-Poisson system, a coupled, nonlinear high-dimensional PDE system used in kinetic plasma simulations as well as in beam physics. Particle-based methods, specifically particle-in-cell (PIC) \cite{birdsallPlasmaPhysicsComputer2018, hockneyComputerSimulationUsing1988}, are one of the most common choices for solving this system. There are different flavors of PIC schemes, each having their own strengths depending on the simulation domains and constraints. They mainly differ in the way the particle-grid interpolations and field solves are performed. Recently, novel particle-field schemes have been developed, such as Particle-in-Fourier (PIF) \cite{mitchell2019efficient,shen2024particle,muralikrishnan2025parapif}. PIF retains the structure and simplicity of standard PIC schemes but offers significantly better conservation and stability properties. Given the diversity of the current HPC landscape it is not clear from an application perspective which of these schemes should be chosen to perform efficient simulations.

The main contribution of this work is to perform a comparison of different flavours of PIC schemes as well as the PIF scheme, all implemented in the performance portable C++ library IPPL \cite{matthias_frey_2024_10878166}, and characterize their scalability and performance on massively parallel supercomputers. Such a study for the electrostatic Vlasov-Poisson system considering the full PIC loop would greatly benefit the beam physics and kinetic plasma communities. Similar studies, albeit only for the field solve on CPUs, include \cite{gholamiFFTFMMMultigrid2016}, where they compare tree-methods such as FMM with FFT and FEM for the Poisson solve, and \cite{bourneSolverComparisonPoissonlike2023}, which compares solvers for a 2D Poisson's equation in the context of gyrokinetic codes.

The comparison is performed on the PIC and PIF implementations in the IPPL library. IPPL is a C++ library providing performance portable and dimension-independent building blocks for scientific simulations which require particle-mesh methods. To deliver this massively parallel and portable particle-mesh toolkit, IPPL makes use of Kokkos \cite{carteredwardsKokkosEnablingManycore2014}, HeFFTe \cite{ayala2020heffte}, and MPI (Message Passing Interface) for inter-node parallelism. IPPL has a modular framework allowing users to mix-and-match available components such as fields, particles, solvers, interpolations, domain decomposition, load balancing, etc. for their particle-mesh simulations, without being bound to a specific hardware. This modularity can be exploited to develop novel numerical algorithms while inheriting performance portability from the framework, as shown in \cite{mayaniMassivelyParallelPerformance2024, muralikrishnan2025parapif}. The availability of plasma physics mini-applications \cite{muralikrishnanScalingPerformancePortability2024} with known solutions ensures correctness at every developmental stage, and may be used for performance testing of the full PIC loop. One of IPPL’s important applications is the production level particle accelerator library OPALX \cite{adelmannOPALXFramework2025, adelmannOPALVersatileTool2019}, which uses IPPL for all its numerical and HPC components.

The paper is structured as follows: In \Cref{theory-pic}, we give a brief overview of the Particle-in-Cell method, as well as the test case which we use for benchmarking. Various numerical solvers for PIC, as well the PIF scheme, are introduced in \Cref{methods}. A comprehensive evaluation of our implementation through correctness and scaling studies on multiple architectures to showcase the portability and performance is given in \Cref{results}. Finally, concluding remarks are stated in \Cref{conclusion}.


\section{Particle-in-Cell} \label{theory-pic}

We introduce the Particle-in-Cell (PIC) scheme, used to simulate the Vlasov-Poisson system. We focus on the dynamics of electrons (mass $m_e$ and charge $q_e$) inside a neutralizing ion background (such that the total charge of the system is 0), and with a fixed external magnetic field $\vec{B}_{ext}$. The evolution of the electron distribution function $f(\vec{x}, \vec{v})$ is described by

\begin{equation}\label{vlasov}
    \frac{\partial f}{\partial t} + \vec{v}\cdot\nabla_{\vec{x}} f + \frac{q_e}{m_e}(\vec{E} + \vec{v}\times\vec{B}_{ext})\cdot\nabla_{\vec{v}}f = 0, 
\end{equation}
where $\vec{E} = \vec{E}_{int} + \vec{E}_{ext}$ is the sum of the electric field due to the Coulomb interaction between the electrons $\vec{E}_{int}$ and the external electric field $\vec{E}_{ext}$. The electric field induced by the Coulomb interaction of the electrons is obtained by solving Poisson's equation:
\begin{equation}\label{poisson}
   \Delta \phi = - \rho /\epsilon_0 \text{ and }   \vec{E}_{int} = -\nabla \phi, 
\end{equation}
where the charge density $\rho = \rho_e + \rho_i$ is the charge density $\rho_e$ of the electrons plus the charge density $\rho_i$ of the neutralizing ion background, $\epsilon_0$ is the vacuum permittivity, and $\phi$ is the electric potential. The charge density is calculated as $\rho_e(\vec{x}) = q_e\int f d\vec{v}$, and the total charge of the electrons is given by $Q_e = \int\rho_ed\vec{x}$. The neutralizing ion background is therefore of total charge $-Q_e$, and has charge density $\rho_i = -Q_e/\int d\vec{x}$.

The Particle-in-Cell algorithm discretizes this system using macro-particles in Lagrangian space, i.e. we store their positions $\vec{x}$ and velocities $\vec{v}$. In order to move to the next time step, instead of solving for the force between each pair of macro-particles, we move to an Eulerian framework. This is done by interpolating from the particle positions onto a grid on the computational domain using some chosen shape functions \cite{saezParticleinCellAlgorithmsPlasma2011}. On the grid, we solve Poisson's equation to obtain $\vec{E}_{int}$, and interpolate back to the particle positions using the same shape functions as before. Finally, we push each particle according to the non-relativistic equations of motion corresponding to \Cref{vlasov}:
\begin{align}\label{eom}
    \frac{d\vec{x}}{dt} & = \vec{v}, \\
    \frac{d\vec{v}}{dt} & = \frac{q_e}{m_e}\left(\vec{E}_{int} + \vec{E}_{ext} + \vec{v}\times \vec{B}_{ext}\right).
\end{align}

\begin{figure}[h]
    \centering
    \resizebox{\linewidth}{!}{
        \begin{tikzpicture}[
            node distance=4ex and 0em,
            block/.style={rectangle, draw=blue!60, fill=blue!5, 
                text width=8em, text centered, rounded corners, minimum height=3em},
            block1/.style={rectangle, draw=green!60, fill=green!5, 
                text width=9em, text centered, rounded corners, minimum height=3em},
            line/.style={draw, -latex},
            ]
            \node [block] (1) {\textbf{SOLVE:} \\ Calculate $\vec{E}_{int}$ from $\rho$};
            \node [block, below right= of 1] (2) {\textbf{GATHER:} \\ Interpolate fields from grid to particles};
            \node [block, below left= of 2] (3) {\textbf{PUSH:} \\ Update particle position and velocity  };
            \node [block, above left= of 3] (4) {\textbf{SCATTER:} \\ Interpolate charge of macro-particles onto grid to obtain $\rho$};
            \node [block1, left=1cm of 4] (5) {\textbf{INITIALIZATION:} \\ Initialize particle positions, velocities, and charges};
            
            \path [line] (5.east) to[out=0, in=180] (4.west);
            \path [line] (1.east) to[out=0, in=90] (2.60);
            \path [line] (2.-60) to[out=-90, in=0] (3.east);
            \path [line] (3.west) to[out=180, in=-90] (4.-120);
            \path [line] (4.120) to[out=90, in=180] (1.west);
    \end{tikzpicture}}
    \caption{The PIC loop starts with initialization of the particle positions and velocities by sampling from an initial distribution function, and then goes into time stepping. At each time step, the charge density of the particles is computed on the grid points of a fixed mesh (\textbf{scatter}), on which we solve for the electrostatic fields (\textbf{field solve}). These are then interpolated to the macro-particle locations (\textbf{gather}) in order to compute the force which will drive the particles to their new position at the next time step (\textbf{particle push}).}
    \label{pic-loop}
\end{figure}
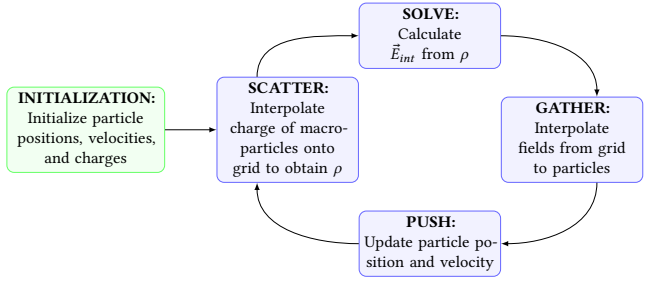

We repeat this until we have simulated the desired amount of time steps, in a process known as the PIC loop, shown schematically in \Cref{pic-loop}. 

\subsection{Landau Damping}

Landau Damping is a well-known plasma physics problem with readily available analytical results, making it a useful mini-application. We consider the following initial electron distribution \cite{ricketsonSparseGridTechniques2016}:
\begin{multline}
    f(t=0) = \frac{1}{(2\pi)^{3/2}}e^{-|\vec{v}|^2/2}(1+\alpha\text{cos}(kx)) \\ \cdot(1+\alpha\text{cos}(ky))\cdot(1+\alpha\text{cos}(kz)), 
\end{multline}
where the size of the domain is $[0,L]^3$ with $L=\frac{2\pi}{k}$, and we choose $k=0.5$ and $\alpha=0.05$ (weak Landau damping). The boundary conditions are periodic in all the dimensions. Even though some of the solvers we use can be applicable to more general boundary conditions, we specifically select this setting for simplicity such that all presented methods can be employed. This is the starting point of the system described by \Cref{vlasov}, and it is  evolved using the PIC scheme (\Cref{pic-loop}).

We use the same set-up as in \cite{muralikrishnanScalingPerformancePortability2024} to check the correctness of our simulations, as well as to perform the scaling studies.

\section{Methods} \label{methods}
The PIC loop has a solve step which requires solving Poisson's equation (\Cref{poisson}). For this purpose, many numerical solvers are available, each suited to different use cases. The ones available in the IPPL codebase are an FFT solver for periodic boundaries, used in \cite{muralikrishnanScalingPerformancePortability2024}, an FFT solver for free-space (\cite{mayaniMassivelyParallelPerformance2024}), a matrix-free finite differences based Preconditioned Conjugate Gradient (PCG) solver, and a matrix-free Finite Element Method (FEM) solver. We use matrix-free formulations of the finite difference and finite element methods to avoid explicit matrix assembly on GPUs, where memory capacity is a limiting factor. For the FFT-based solvers, we rely on the external library \texttt{heFFTe} \cite{ayala2020heffte}, which is specifically designed for exascale and heterogeneous computing platforms. Furthermore, in IPPL, we also have the Particle-in-Fourier (PIF) scheme \cite{muralikrishnan2025parapif}, which follows the idea of PIC but avoids interpolation onto a real space grid by directly projecting particle charges onto the Fourier space, perform field solve on the Fourier space and finally interpolate the electric field from the Fourier space directly to the particle locations.

The PIF scheme requires non-uniform FFTs for the projections between the particles and the Fourier space. Since \texttt{heFFTe} supports only uniform FFTs, we implement our own non-uniform FFT routine in IPPL which utilizes \texttt{heFFTe} for uniform FFTs in its sub-steps.

All the methods presented have different strengths and weaknesses, which makes the choice of solver for a simulation dependant on the user's requirements. A summary table of solver properties to aid such a decision can be found in \Cref{solver_comparison_table}.

\begin{table}[h]
\centering
\begin{tabular}{| c | >{\centering\arraybackslash}m{5.5em}  >{\centering\arraybackslash}m{4.8em}  >{\centering\arraybackslash}m{8em} |}
\hline
\textbf{Solver} & \textbf{Grid uniformity} & \textbf{Boundary conditions} & \textbf{Accuracy} \\ \hline
\textbf{FFT} & Required        & Periodic, Free space                     & Pseudo-spectral \\ \hline
\textbf{PCG} & Required        & Periodic, Dirichlet, Neumann          & 2nd order \\ \hline
\textbf{FEM} & Not required    & Periodic, Dirichlet, Neumann          & $(p+1)$ order where $p$ is the polynomial order of the basis functions \\ \hline
\textbf{PIF} & Not required    & Periodic, Free space                            & Spectral \\ \hline
\end{tabular}%
\caption{Summary of main solver properties which influence the choice of solver for a simulation.}
\label{solver_comparison_table}
\end{table}

\subsection{FFT solver}

In the case of periodic boundary conditions, one can solve Poisson's equation in Fourier space easily, since derivatives become multiplication by the wave-vector in Fourier space. The solution of \Cref{poisson} can be written as
\begin{equation*}
    \phi = \mathcal{F}^{-1} \left( \frac{1}{k^2} \mathcal{F}\left(\frac{\rho}{\epsilon_0}\right)\right) \text{ and } \vec{E}_{int} = \mathcal{F}^{-1} \left( -i\vec{k}\frac{1}{k^2} \mathcal{F}\left(\frac{\rho}{\epsilon_0}\right)\right),
\end{equation*}
where $\mathcal{F}(\cdot)$ represents the Fourier transform of a function, $\vec{k}\in\mathbb{R}^3$ is the wave-vector, and the imaginary unit is defined as $i=\sqrt{-1}$. Using an FFT-based grid solve for PIC is beneficial because the accuracy is only dependent on the source $\rho$, and not the numerical accuracy of the discretization as in finite differences. However, the drawback is that one is limited to periodic or free space boundary conditions when using this solver. We use the \texttt{heFFTe} library for the FFTs in IPPL. Information on the computational implementation and scaling properties of this library can be found in \cite{ayala2020heffte}.
 
\subsection{PCG solver}

We discretize the Laplace operator in \Cref{poisson} using second order central finite differences. This results in the linear system $A\vec{x} = \vec{b}$, where $A$ is the Poisson matrix given by the finite difference discretization of the Laplace operator, $\vec{x}$ is the solution vector $\phi$ at the discrete grid-points, and $\vec{b}$ the source term $\rho$ at the discrete grid-points. Boundary conditions are also taken care of in the matrix $A$. We solve the system $A\vec{x} = \vec{b}$ iteratively, using the Conjugate Gradient (CG) algorithm with a matrix-free version of $A$, i.e., we only provide the CG algorithm with the action of the matrix on a vector instead of storing the whole matrix. We have a suite of preconditioners available for the CG algorithm in IPPL, such as Jacobi, Richardson, Gauss-Seidel, and Symmetric Successive Over-Relaxation (SSOR). The advantages of this scheme are its simplicity, completely local communication and ability to handle Dirichlet and Neumann boundary conditions in addition to periodic boundaries. However, it is only second order accurate and also limited to Cartesian domains with uniform mesh spacing in each dimension. 

\subsection{The Finite Element Solver}

The Finite Element Method (FEM) is widely used to solve Poisson's equation in many areas of physics. In PIC, it is commonly used due to its ability to handle unstructured meshes, provide higher-order accuracy and to simulate electromagnetic phenomena \cite{crawfordHigherOrderCharge2021, hiptmairFiniteElementsComputational2002}. FEM also reduces to a linear system of equations $A\vec{x} = \vec{b}$ which can be solved using an iterative solver like the CG algorithm if $A$ is symmetric positive definite. The drawback of building the entire FEM matrix $A$ can be avoided by implementing matrix-free FEM. This is crucial for GPU usage, as storing and retrieving a full finite element matrix $A$ on GPUs is a costly endeavour, especially at higher orders. With this idea, IPPL supports matrix-free FEM PDE solvers, which use a functor representing the action of the matrix $A$ for a specific PDE on a vector $\vec{x}$ as the operator passed to the CG method. The CG method is then used to solve the finite element system iteratively.

In the case of Poisson's equation, for a domain discretized with $N_e$ elements and containing $N_{dofs}$ degrees of freedom, the global FEM matrix $A = \sum_{e=1}^{N_E} A^e$ will be given by
\begin{equation*}
    \sum_{e=1}^{N_E} \sum_i \phi_i \underbrace{\int_{\Omega_e} d\Omega_e  \nabla b_i \cdot \nabla b_j}_{=: A^e_{i,j}} = \sum_{e=1}^{N_E} \int_{\Omega_e} d\Omega_e \rho \cdot b_j,
\end{equation*}        
where $\Omega_e$ is the $e$-th element in the mesh, $b_i$ are the basis functions of our Finite Element space, with $i=1, ..., N_{dofs}$, $\phi_i$ is our solution at the degrees of freedom, and $\rho$ is the source term (right-hand side of Poisson's equation). For a derivation and introduction to FEM, we refer the reader to \cite{hiptmairNumericalMethodsPartial2024}.

For the Poisson solver comparison, we will use first-order Lagrange basis functions, which give second-order accuracy. Implementation of higher-order Finite Element spaces is still an ongoing project in IPPL.

As with the CG-based finite differences solver (which we call PCG above), we can also apply preconditioning to the CG algorithm used with the Finite Element Method. This is supported by the IPPL library. However, matrix-free preconditioning is non-trivial when using FEM, since we do not always know the analytical form of the matrix, a side effect of the general way in which it is implemented. For the finite differences case, matrix-free preconditioning for the Laplace operator is easier since one knows the analytical form of the matrix $A$, and can therefore implement the action of different matrix decompositions which are required for preconditioning, such as upper, lower, diagonal, etc. In \Cref{results}, we will therefore only compare with plain CG + FEM, and discuss preconditioning separately in \Cref{preconditioning_fem}.

\subsection{Particle-in-Fourier (PIF)}

The Particle--in--Fourier (PIF) approach provides an alternate mechanism to PIC for coupling Lagrangian particle data directly to a spectral field representation. In PIF, each particle carries a weight \(q_j\) (for example charge) located at position \(\mathbf{x}_j\). Instead of depositing these quantities onto an intermediate mesh, the spectral coefficients are obtained directly through a type~1 non-uniform fast Fourier transform (NUFFT) for computing \(\hat{\rho}(\mathbf{k})\) efficiently,
\[
\hat{\rho}(\mathbf{k}) 
= \sum_{j=1}^{N_p} q_j \, e^{-i\,\mathbf{k}\cdot \mathbf{x}_j},
\qquad \mathbf{k} \in K_N ,
\]
which yields a Fourier-space representation of the particle density \(\rho\). Field equations which are diagonal in Fourier space, such as Poisson's equation
\[
-\|\mathbf{k}\|^2 \, \hat{\phi}(\mathbf{k}) = \frac{1}{\epsilon_0} \hat{\rho}(\mathbf{k}),
\]
can then be solved mode-by-mode. The resulting force field is given spectrally by
\[
\hat{\mathbf{E}}_{int}(\mathbf{k}) = -i\,\mathbf{k}\,\hat{\phi}(\mathbf{k}),
\]
from which physical-space forces at particle positions are recovered via a type~2 NUFFT,
\[
\hat{\mathbf{E}}_{int}(\mathbf{x}_j)
= \sum_{\mathbf{k} \in K_N} 
\hat{\mathbf{E}}_{int}(\mathbf{k}) \, e^{i\,\mathbf{k}\cdot \mathbf{x}_j},
\qquad j = 1,\ldots,N_p .
\]
We refer the readers to \Cref{appendix:nufft} for a more detailed explanation of NUFFTs.
This direct particle--to--Fourier and Fourier--to--particle coupling avoids mesh interpolation, which in turn eliminates aliasing, provides excellent conservation and stability properties, and preserves the spectral accuracy of Fourier solvers. As a result, PIF methods provide a high-fidelity alternative to grid-based particle--mesh algorithms in kinetic plasma simulations and related particle-based models. Apart from the electrostatic context, PIF schemes can also be applied to gyrokinetic \cite{ohana2016towards} and electromagnetic kinetic plasma simulations \cite{shadwick2014variational,ameres2018stochastic,campos2024variational}. While originally limited to periodic dimensions, in \cite{shen2024particle} they have been extended to free space and Dirichlet boundary conditions in certain types of geometries.


\section{Results} \label{results}

\begin{figure}[H]
    \centering
    \includegraphics[width=\linewidth]{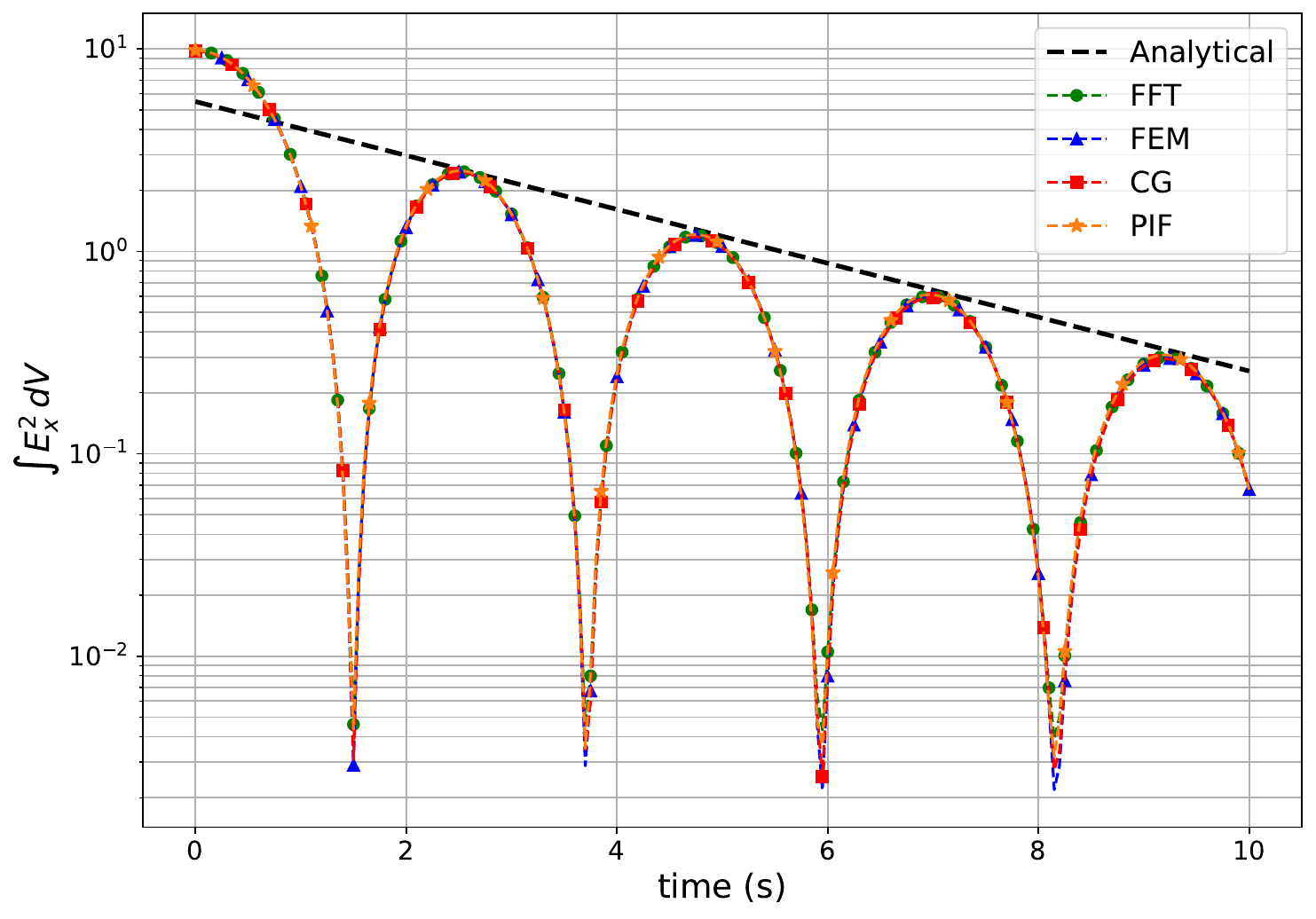}
    \caption{Energy of the electric field component in $x$-direction ($E_x$) for the Landau damping simulation over time. The FFT, FEM, PCG and PIF schemes yield the same solution, and the oscillation peaks exhibit the expected analytical damping rate.}
    \label{correctness_landau}
\end{figure}

We present a comparative performance analysis of the four schemes introduced above, using Landau damping as the benchmark application. The underlying physics of the test case is described in \Cref{theory-pic}. The tolerance of the iterative solvers in PCG and FEM, as well as the NUFFTs in the PIF scheme, are set to be $10^{-4}$, as with this tolerance we still observe the correct analytical damping rate in the simulations. The correctness study using all methods introduced in \Cref{methods} is shown in \Cref{correctness_landau}.

\subsection{Setup}

We perform strong scaling studies for two problem sizes, case A and case B, shown in \Cref{setup_sizes}, as done in \cite{muralikrishnanScalingPerformancePortability2024}. The choice of 8 particles per cell is standard for particle accelerator simulations using PIC \cite{MYERS2021102833}. In other electrostatic PIC applications, where coarser grid sizes and much higher number of particles per cell ($\mathcal{O}(100)-\mathcal{O}(1000)$) are used, the particle operations will dominate the total run time and hence the influence of field solver is less pronounced \cite{muralikrishnanScalingPerformancePortability2024}. We run the simulation for 10 time steps, and consider the total time as the runtime of the simulation minus the initialization time.

\begin{table}[h]
    \centering
    \begin{tabular}{|c|c|c|}
        \hline
        Case & Grid size & No. of particles \\
        \hline
        A & $512^3$ & 1'073'741'824 \\
        B & $1024^3$ & 8'589'934'592 \\
        \hline
    \end{tabular}
    \caption{The two cases of grid sizes and particle counts used in the Landau damping strong-scaling study.}
    \label{setup_sizes}
\end{table}

To assess portability and characterize performance across architectures, we conduct the scaling studies on three different machines:
\begin{itemize}
    \item Daint at Alps (CSCS, Switzerland). This machine has 2688 nodes of 4 sockets each with GH200 (Nvidia Grace-Hopper) chips. 
    \item LUMI-G partition at LUMI (CSC, Finland), which has 2978 nodes with 4 AMD MI250X GPUs of two compute dies per node, resulting in virtually 8 GPUs per node.
    \item JUWELS Booster at Jülich Supercomputing Center, Germany. This machine has 936 nodes with 4 A100 (Nvidia) GPUs each.
\end{itemize}

For all cases, we run on GPUs with one MPI rank per GPU and utilizing full nodes exclusively. The minimum node count for each scaling study is determined by the memory requirements of the problem size as well as the scheme on the respective machine. For the PCG solver, we use an SSOR preconditioner with four inner and two outer iterations, and a damping factor of $\pi/2$. For all CG-based solvers, we take the solution from the previous time step as our starting guess for the CG iterations in the next time step. On the LUMI machine, we were unable to run some combinations of node counts and problem sizes due to unknown memory issues after the LUMI update in January 2026, which are currently being investigated.

\subsection{Scaling Studies}

\begin{figure}[h]
    \centering
    \includegraphics[width=\linewidth]{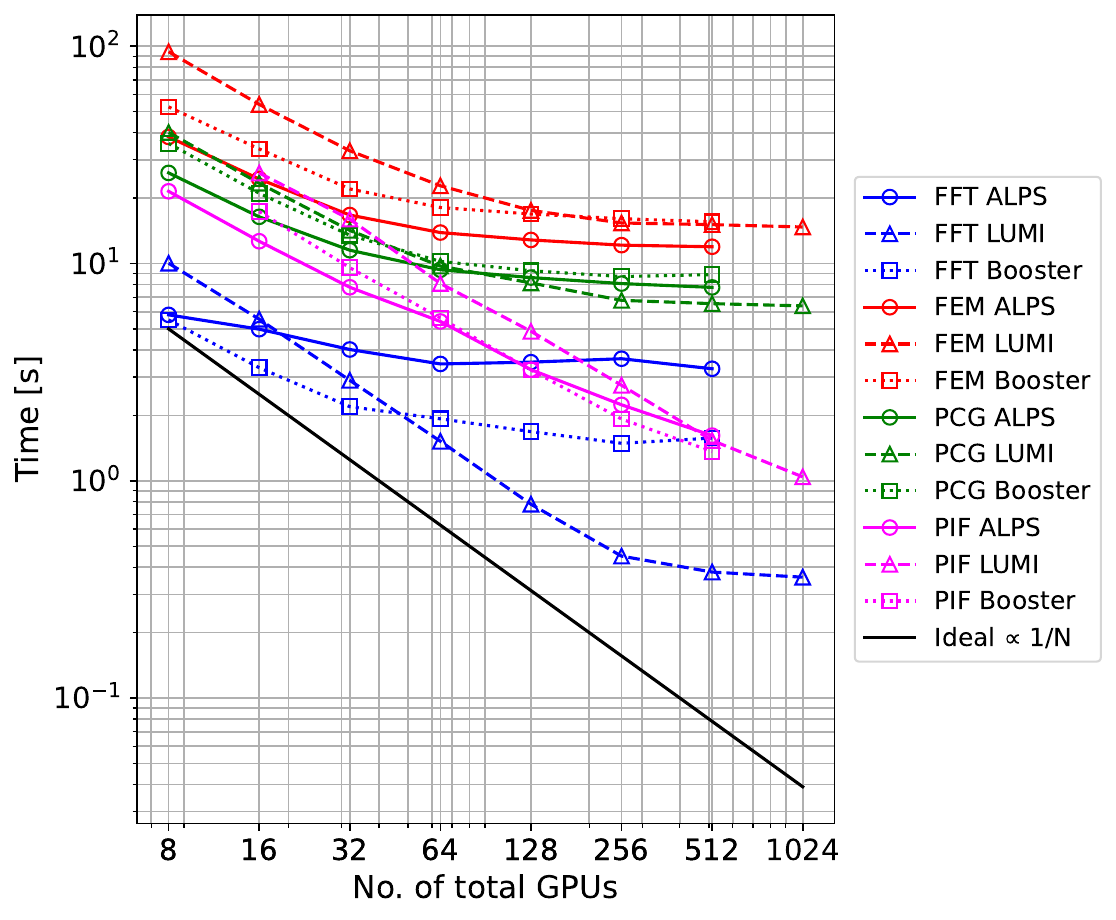}
    \caption{Strong scaling study for case A ($512^3$ and 8 particles per cell) for all presented solvers on all three architectures considered. The black curve represents ideal scaling.}
    \label{scaling_all_512}
\end{figure}

\begin{figure}[h]
    \centering
    \includegraphics[width=\linewidth]{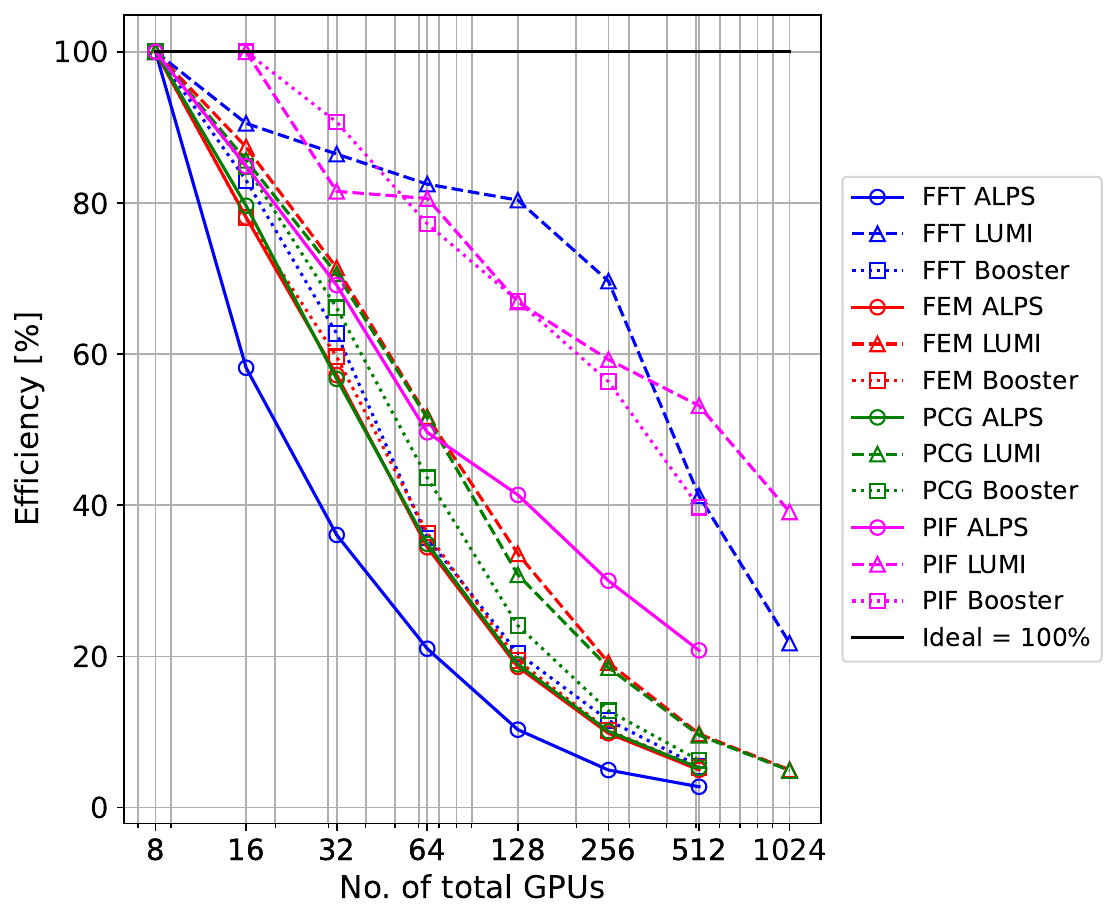}
    \caption{Strong scaling efficiency for case A ($512^3$ and 8 particles per cell) for all presented solvers on all three architectures considered.}
    \label{efficiency_all_512}
\end{figure}

\begin{figure}[h]
    \centering
    \includegraphics[width=\linewidth]{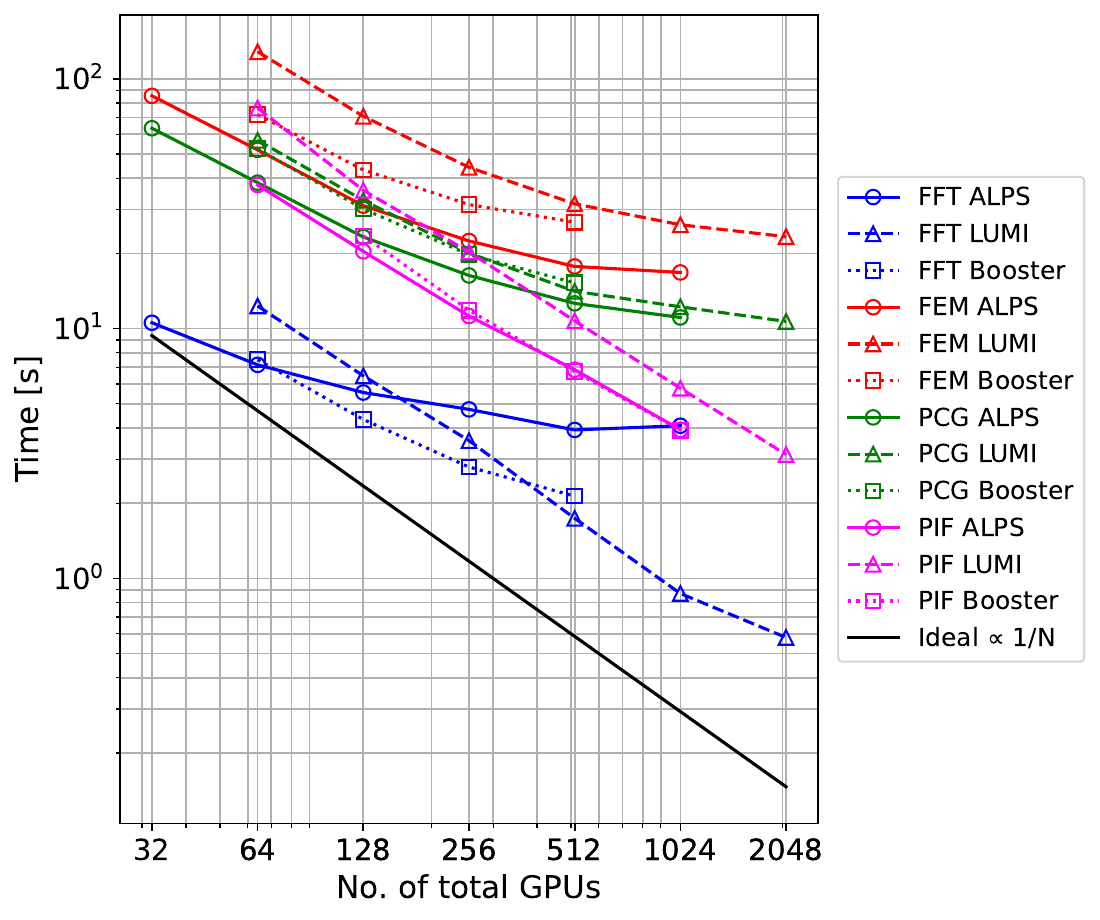}
    \caption{Strong scaling study for case B ($1024^3$ and 8 particles per cell) across all solvers and all three architectures evaluated.}
    \label{scaling_all_1024}
\end{figure}

\begin{figure}[h]
    \centering
    \includegraphics[width=\linewidth]{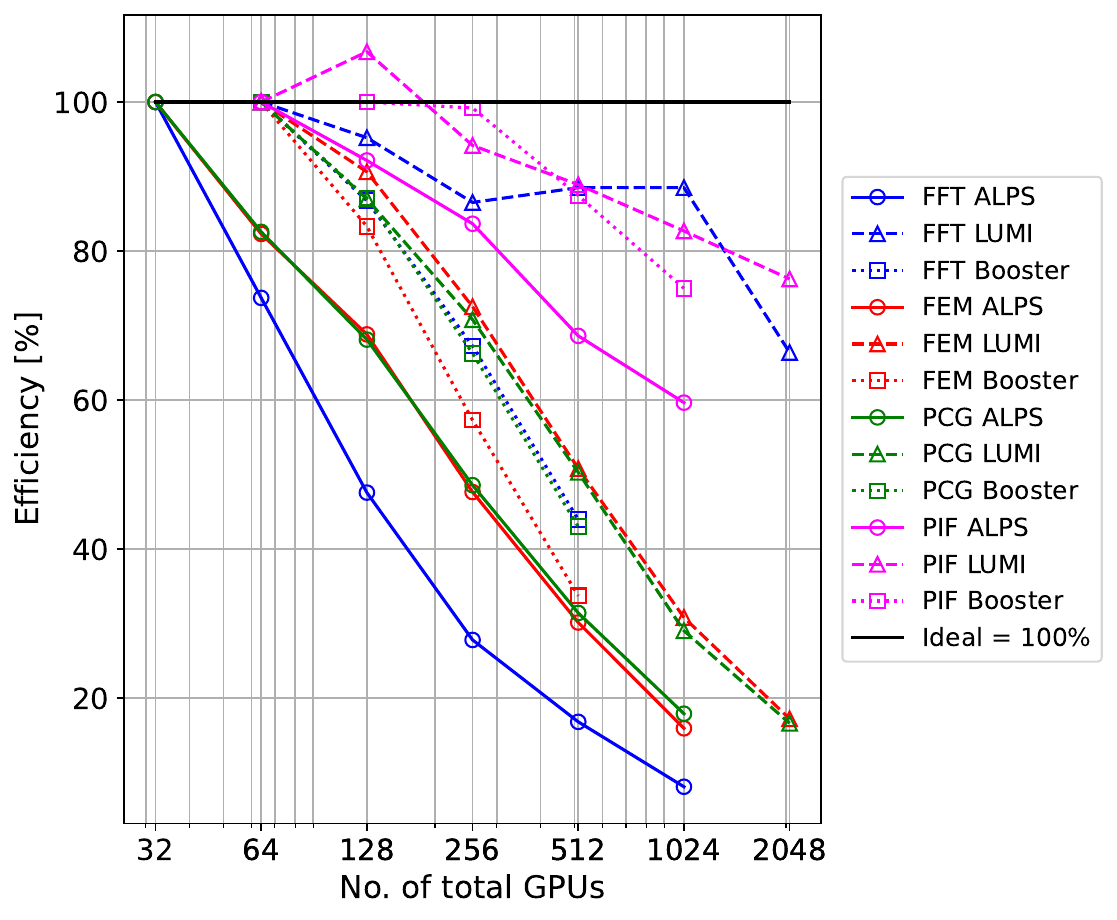}
    \caption{Strong scaling efficiency for case B ($1024^3$ and 8 particles per cell) for all presented solvers on all three architectures considered.}
    \label{efficiency_all_1024}
\end{figure}

The strong scaling results for case A are shown in \Cref{scaling_all_512}, with a comparison of the different solvers on the different machines. \Cref{scaling_all_1024} shows the same for case B. \Cref{efficiency_all_512} and \Cref{efficiency_all_1024} show the strong scaling efficiency for case A and case B, respectively. Efficiencies are computed using the smallest node count on which the problem fits as the baseline.

\subsubsection{Particle update kernel on Alps} 
\begin{figure}[h]
    \centering
    \includegraphics[width=\linewidth]{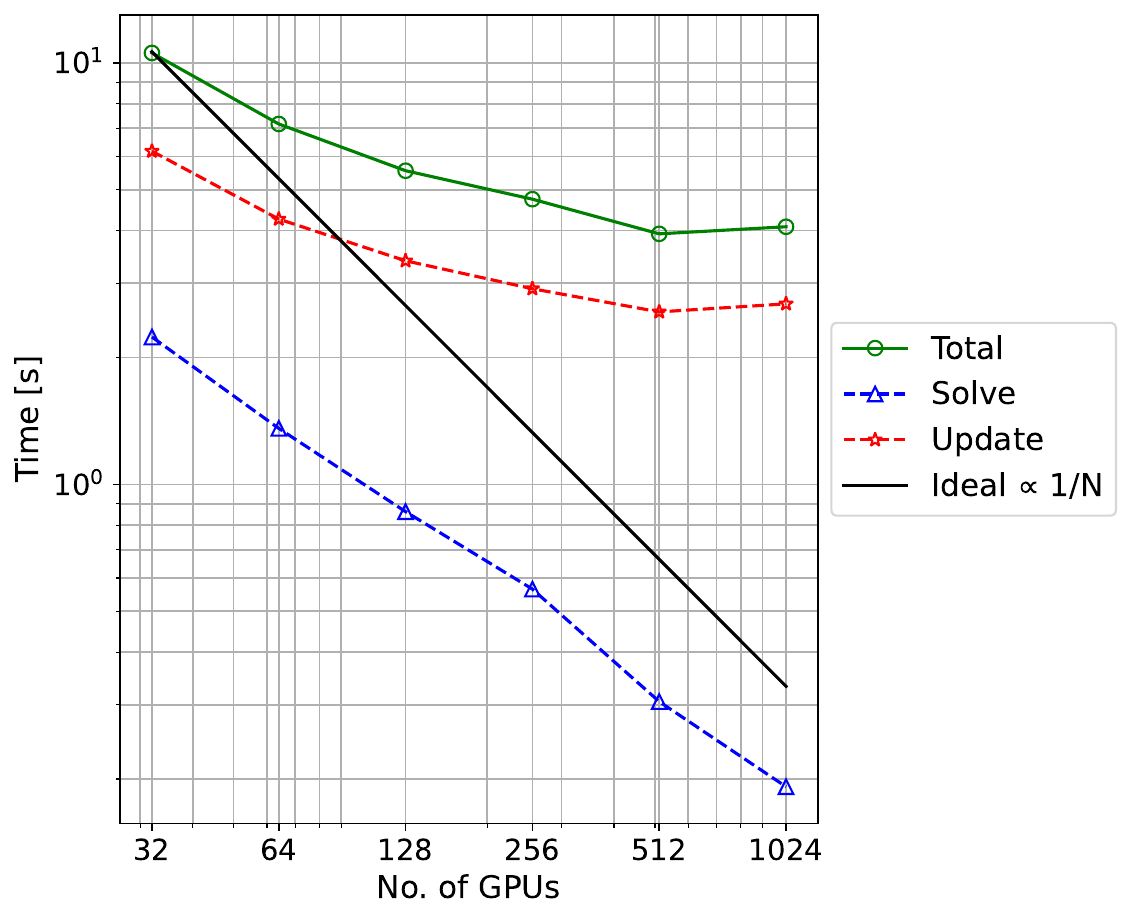}
    \caption{Breakdown of the total time of the Landau Damping simulation with the FFT solver for case B into the dominant kernels, on Alps. These are the `solve' kernel (where the FFT is used to solve Poisson's equation on the grid), and the `update' kernel, where the particles are sent to their new position after the particle push (and communicated among ranks if necessary).}
    \label{FFT_1024_alps}
\end{figure}

On LUMI and Booster, for both problem sizes the FFT solver is the best method in terms of absolute time. This goes to show that it is beneficial to use the external \texttt{heFFTe} library, targeted specifically towards massively parallel and portable workloads, as we reap the benefits of their optimizations. On Alps, the FFT computation itself scales well. However, enabling CUDA IPC on GH200 increases the absolute runtime of the particle-update kernel in the PIC loop and prevents it from scaling (see \Cref{FFT_1024_alps}). Because the FFT phase is both fast and well-scaling, the particle-update kernel becomes the dominant bottleneck in the FFT-solver configuration. In contrast, for the PCG, FEM, and PIF solvers—where the primary bottlenecks lie in the linear solve (PCG and FEM) or in the scatter/gather operations (PIF)—the particle-update cost is largely hidden, leading to significantly better scaling on Alps.

\subsubsection{Strong scaling efficiency on different machines} For the small problem size, i.e. case A, the scaling efficiency on Alps drops below 50\% after 32 GPUs when using the FFT solver, illustrating the strong impact of the CUDA IPC issue on the particle-update kernel. For the FEM, PCG, and PIF solvers, this drop occurs after 64 GPUs. 

On the other hand, on the JUWELS Booster machine with the older Nvidia A100 architecture, the efficiency for FFT, FEM, and PCG solvers drops below 50\% after 64 GPUs. The FFT solver scales better on Booster than on Alps. This could be an interplay of two effects: the optimized FFT kernels having significantly lower absolute runtimes on the newer Grace–Hopper architecture, which reduces available parallel work and therefore weakens strong scaling behaviour; and the Booster having a better interconnect for MPI communication than Alps. 

For the PIF solver, the efficiency remains above 50\% up to 512 GPUs and 256 GPUs on LUMI and JUWELS Booster, respectively. A similar behaviour is observed for the FFT solver on LUMI, where the 50 \% threshold is crossed only beyond 512 GPUs. For the FEM and PCG solvers on LUMI and JUWELS Booster, the efficiency drops below 50 \% after 64 GPUs.

Since for the PCG and FFT solvers the absolute times on LUMI are comparable or better than on Alps, LUMI seems to be better suited for these solvers in our setup. In contrast, the FEM solver achieves the best absolute performance on Alps, suggesting that this solver benefits from the increased computational throughput of GH200. 

For the PIF scheme, although Alps provides the best absolute time at small node counts, LUMI and Booster become more favourable once larger node counts are used. Despite the slower A100 GPU, the Booster benefits from its faster interconnect, which becomes decisive at scale. 

When it comes to case B, a bigger problem size, LUMI seems to provide the best efficiency again. The FFT solver remains above 50\% efficiency up to 2048 GPUs, and the PCG and FEM solvers cross the 50\% threshold only at 1024 GPUs. For the JUWELS Booster, FFT, FEM, and PCG solvers drop below 50\% at 512 GPUs. On Alps, this happens much sooner, at 256 GPUs for the FEM and PCG solvers, and at 128 GPUs for the FFT solver, reflecting the persistent impact of the CUDA IPC issue. The PIF solver scales well for this larger problem size on all three machines, with efficiencies staying above 50\% up to the highest node count considered, i.e.\ 1024 GPUs on Alps and Booster, and 2048 GPUs on LUMI, staying closer to 80\% for the latter architecture.

\subsubsection{Bottlenecks in FEM} The FEM and PCG solvers exhibit similar scaling behaviour, albeit with a shift in absolute times. This similarity is expected, as both solvers rely on the CG method as their primary computational kernel. The key difference lies in the evaluation of the matrix-vector product required in the iterative solve. 

For the PCG solver, this consists in a simple finite differences scheme, where we know beforehand the form of the Laplacian matrix. Boundary conditions are automatically enforced by the exchange of ghost cells. 

In the case of the FEM solver, the matrix-vector evaluation is done by using a pre-computed local element matrix (on the reference element) for each element, and accumulating the contributions into the appropriate global vector entries. Boundary degrees of freedom must be treated explicitly to apply boundary conditions correctly. This results in slightly higher computational cost due to the use of conditionals in GPU kernels.

\begin{figure}[h]
    \centering
    \includegraphics[width=0.5\textwidth]{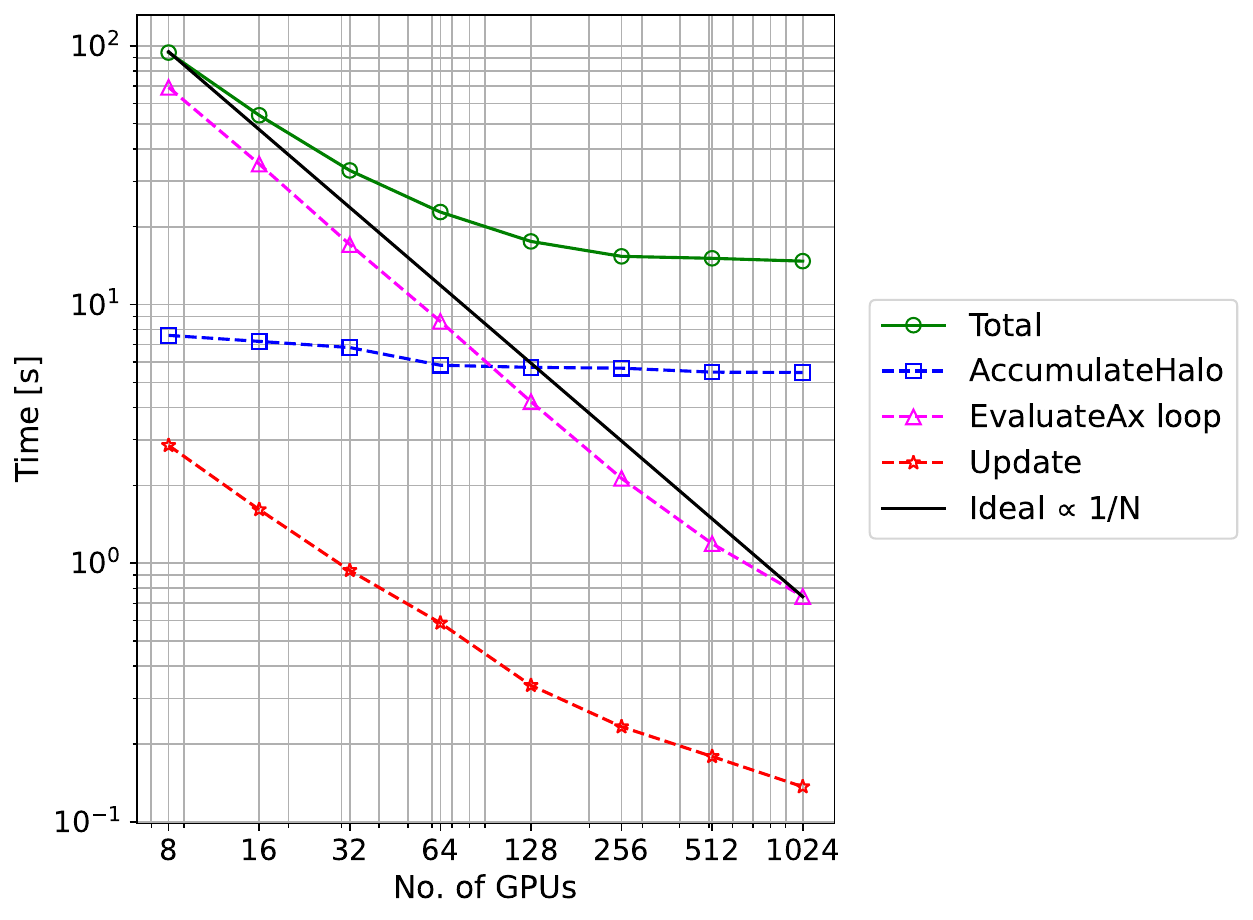}
    \caption{Breakdown of the total time of the Landau Damping simulation using the FEM solver for case A on LUMI. The loop over elements corresponds to `evaluateAx loop', while `accumulateHalo' denotes the communication kernel which limits the scaling. Both kernels dominate the particle update kernel, effectively hiding its cost.}
    \label{FEM_512_lumi}
\end{figure}

Despite the use of atomic operations, the element-wise matrix-vector computation within the Kokkos kernel scales very well from \Cref{FEM_512_lumi}. The main bottleneck for FEM is the halo cell accumulation required for degrees of freedom shared among MPI ranks. This communication step must occur at every iteration in the CG solver, and therefore happens hundreds of times per solve, as can be seen from the iteration counts in \Cref{preconditioner_its}. This is reflected in the high halo communication costs in \Cref{FEM_512_lumi}. One could improve this by overlapping the communication with the computation.

\subsubsection{Weak scaling}

\begin{figure}[h]
    \centering
    \includegraphics[width=\linewidth]{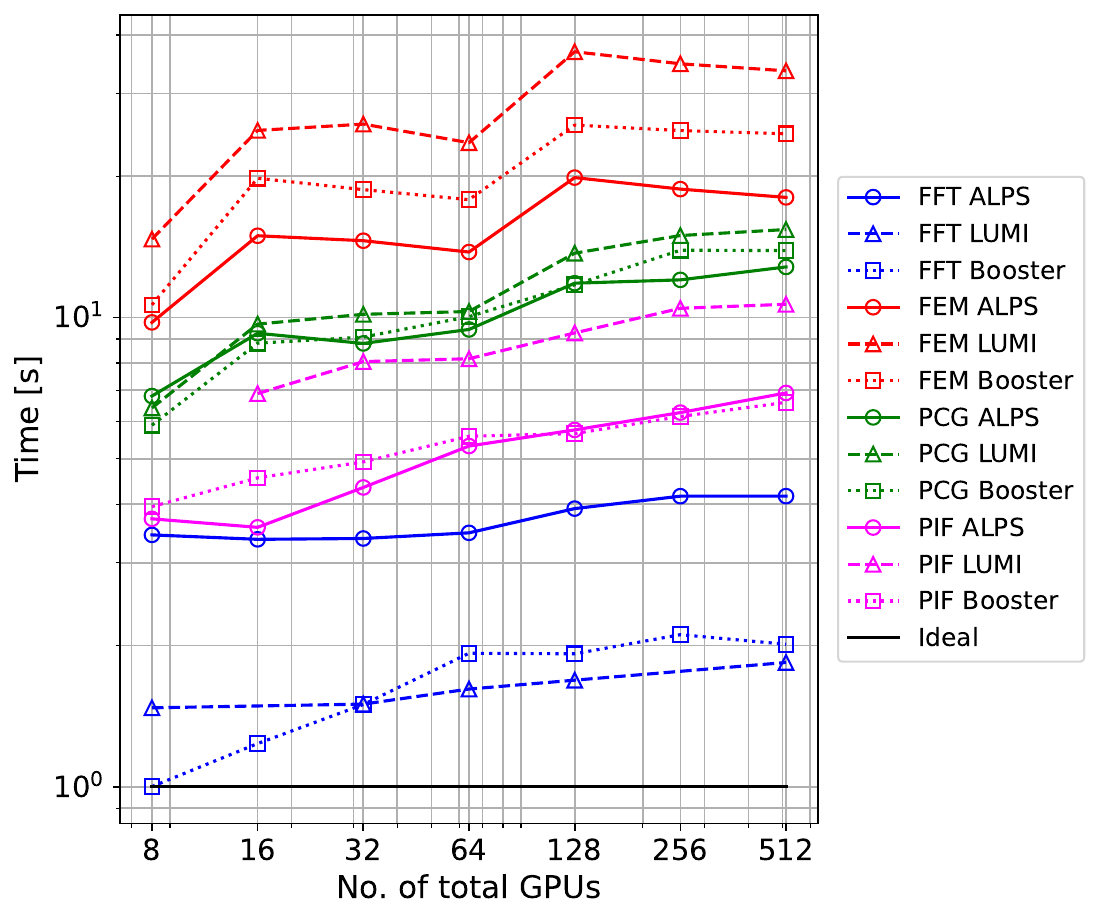}
    \caption{Weak scaling for the four solvers on all architectures. Ideally, the runtime should stay constant as we increase the GPU count.}
    \label{weak_scaling_all}
\end{figure}

\begin{figure}[h]
    \centering
    \includegraphics[width=\linewidth]{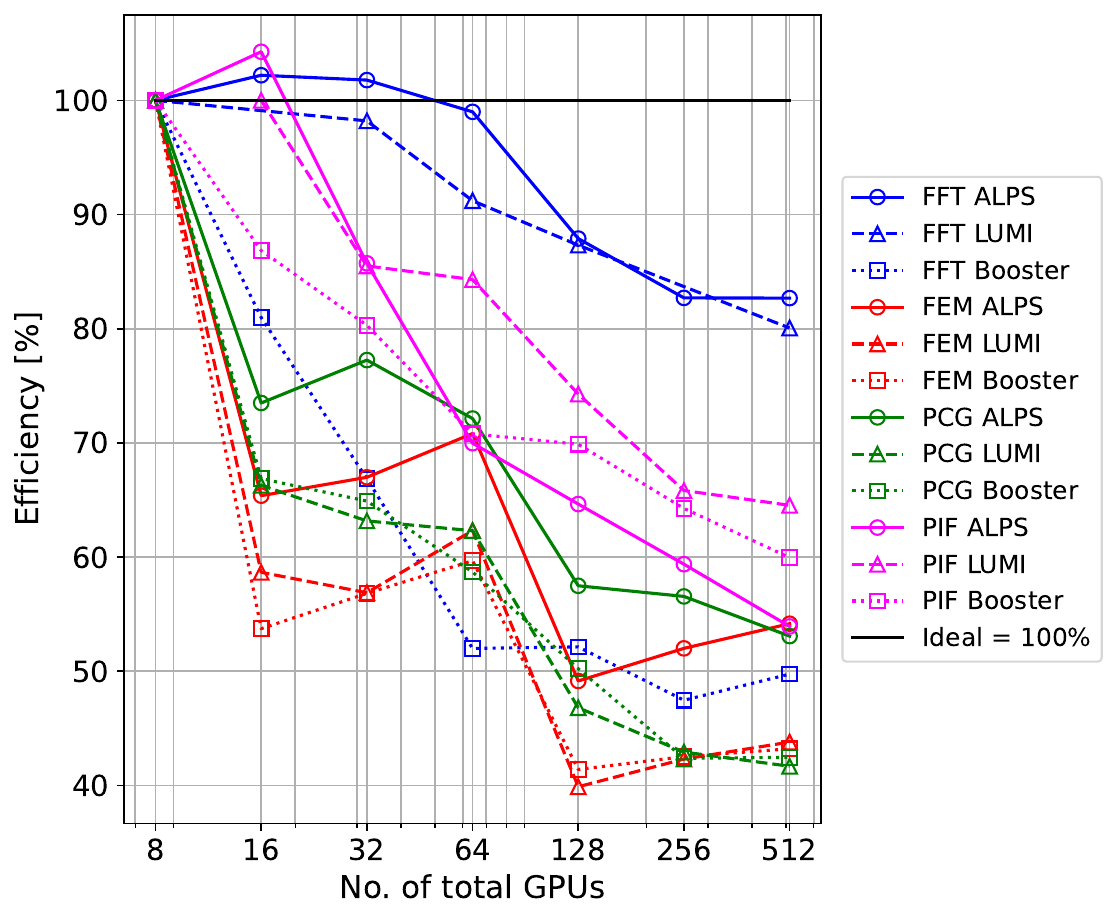}
    \caption{Weak scaling efficiency for the four solvers on all architectures.}
    \label{weak_scaling_efficiency}
\end{figure}

We run weak scaling studies for all solvers on all three machines. For this study, we start with a problem size of $256^3$ and 8 particles per cell on 8 GPUs, and proportionally increase the problem size as we increase the number of GPUs to keep the work per GPU constant. The setup of the simulation parameters can be found in \Cref{weak_scaling_setup}. The resulting scaling studies are presented in \Cref{weak_scaling_all}, with the weak scaling efficiency being shown in \Cref{weak_scaling_efficiency}.

\begin{table}[h]
\centering
\begin{tabular}{|c|c|c|}
\hline
No. of GPUs & Gridsize           & No. of Particles \\ \hline
8           & $256^3$            & 134'217'728      \\ \hline
16          & $256^2 \cdot 512$  & 268'435'456      \\ \hline
32          & $256 \cdot 512^2$  & 536'870'912      \\ \hline
64          & $512^3$            & 1'073'741'824    \\ \hline
128         & $512^2 \cdot 1024$ & 2'147'483'648    \\ \hline
256         & $512 \cdot 1024^2$ & 4'294'967'296    \\ \hline
512         & $1024^3$           & 8'589'934'592    \\ \hline
\end{tabular}%
\caption{Problem size setup for the weak scaling study, which start at $256^3$ and end at $1024^3$, always keeping 8 particles per cell.}
\label{weak_scaling_setup}
\end{table}

The FEM solver and the PCG solver have similar weak scaling behaviours, due to the communication patterns being similar for both. They have a large drop in efficiency which is most probably due to the reductions and communication required at each iteration of the iterative solver.

The FFT solver scales well, staying above 80\% weak scaling efficiency until 512 GPUs on both Alps and LUMI. However, on the JUWELS Booster it drops below 50\% at 256 GPUs. This may be due to the FFT communication pattern behaving differently on JUWELS Booster, since it has a different interconnect (Infiniband) than Alps and LUMI (which both have Slingshot). Nevertheless, it seems to plateau to a constant behaviour after an initial ramp-up phase, as seen in \Cref{weak_scaling_all}.

The PIF solver does not achieve the same scaling efficiency as the FFT solver, but stays above 60\% until 512 GPUs on both LUMI and the JUWELS Booster machine. On Alps, it still stays above 50\% at 512 GPUs. With the work in terms of modes/grid points and number of particles per GPU being constant, the increased communication costs due to the use of wider stencils in the NUFFTs could explain the degradation in weak scaling efficiency of the PIF scheme compared to the pseudo-spectral FFT PIC scheme, which uses the standard Cloud-in-Cell interpolation scheme.

\subsubsection{Applicability of different solvers} Overall, in terms of absolute timings, the pseudo spectral FFT PIC scheme is the clear winner. However, it is not a general purpose method. PIC simulations are used in many contexts, which may require more flexible boundary conditions than naturally supported by FFTs. The strength of the PCG and FEM solvers lies precisely in this flexibility—both in the choice of boundary conditions and, in the FEM case, the ability to operate on non-uniform meshes.

Our results show that the PCG and FEM solvers scale similarly to the FFT solver, even though their absolute runtimes are roughly an order of magnitude higher. The novel PIF solver is interesting in cases where accuracy, conservation and stability properties are essential.  In terms of absolute timings it stays close to the PCG solver at lower node counts, beating the traditional PIC FFT solver at higher node counts due to its more favourable scaling properties. However, this is likely due to the particle update issue in the FFT solver case, which can still be optimized. In any case, this shows that higher order schemes like PIF can take full advantage of massively parallel computing architectures.  

\subsection{Preconditioning for FEM} \label{preconditioning_fem}
\begin{figure}[h]
    \centering
    \includegraphics[width=0.75\linewidth]{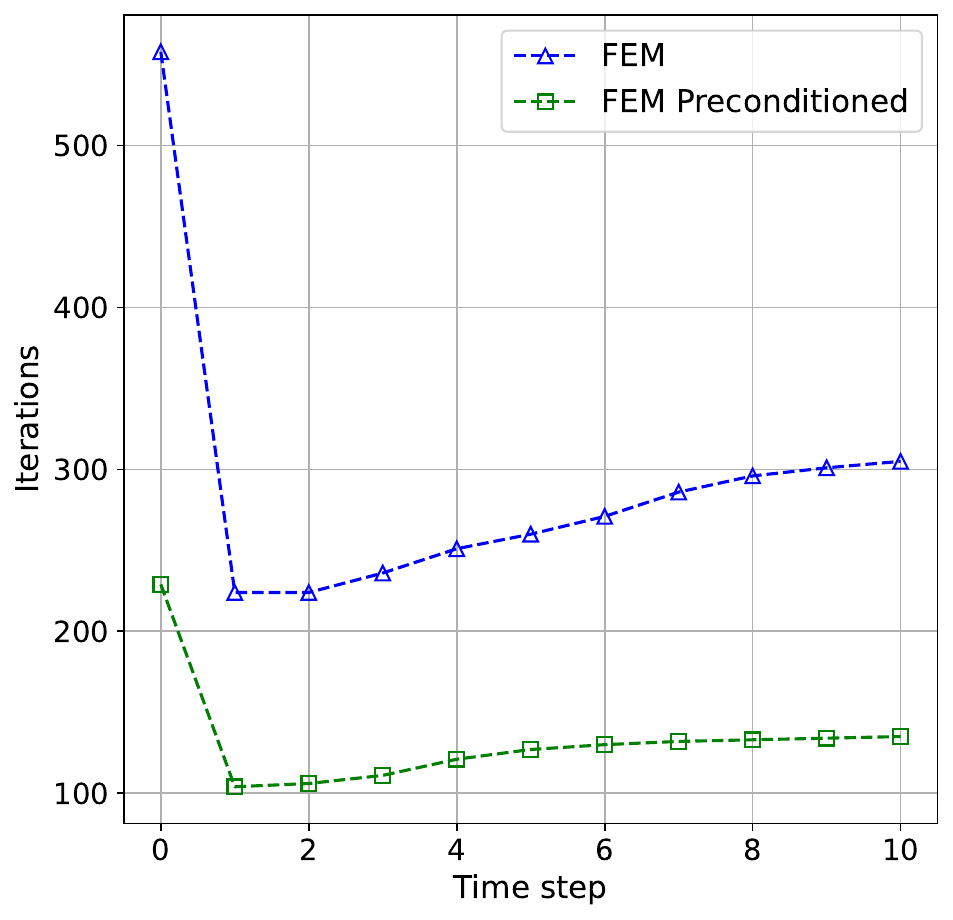}
    \caption{CG iteration counts per time step with the FEM solver for case A on two nodes on Alps, comparing runs with and without preconditioning.}
    \label{preconditioner_its}
\end{figure}
\begin{figure}[h]
    \centering
    \includegraphics[width=0.75\linewidth]{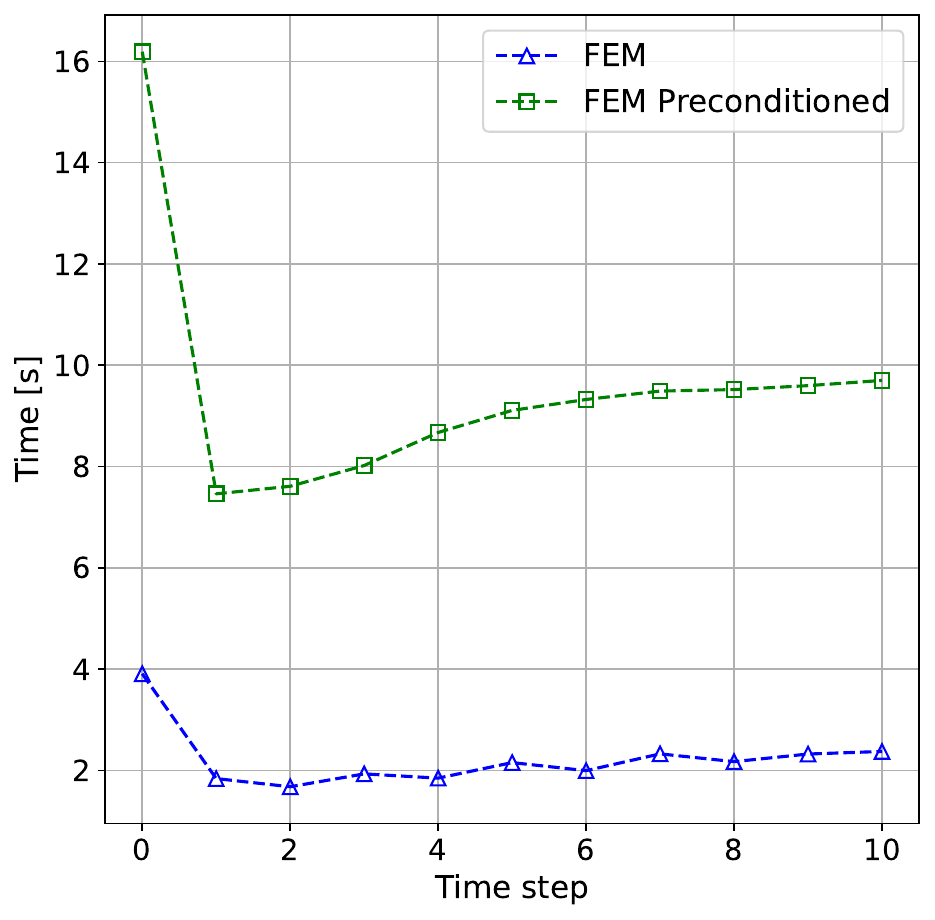}
    \caption{Time spent per time step on the CG iterations for case A on 2 nodes on Alps, for the FEM solver with and without preconditioning.}
    \label{preconditioner_time}
\end{figure}

Matrix-free preconditioning in the case of FEM is non-trivial due to the cost of identifying different matrix decompositions without constructing the matrix. In the finite difference case, due to the known structure of the Laplacian matrix, this can easily be implemented as a matrix-free operator which maps the input vector to the correct output solution. 
In FEM, the absence of the fully assembled operator makes it challenging to distinguish matrix blocks without introducing conditional logic.
A naive approach is to use ``if" statements inside the preconditioner, which is what our current implementation does. The naive preconditioner behaves as expected —it reduces the number of CG iterations required for convergence— but it increases the runtime per iteration due to its computational expense, as shown in \Cref{preconditioner_its} and \Cref{preconditioner_time}. This is due to conditionals on the GPUs being sub-optimal, and remains a weakness in the current FEM implementation.  Our ongoing work is to improve this aspect by exploring matrix-free preconditioning strategies from the literature  \cite{kronbichlerEnhancingDataLocality2023}, with a particular focus on multigrid methods \cite{strangMultigridMethodsMathematical2006}.


\section{Conclusion} \label{conclusion}

In this work, we compare different flavours of Particle-in-Cell schemes which differ in their field solver strategy, as well as the Particle-in-Fourier scheme, each with its own strengths and limitations. Using a performance portable implementation of all the schemes in the open-source, particle-mesh library IPPL, we evaluated their scalability and performance across two different problem sizes and three hardware platforms on the Landau damping plasma physics test case.

For the benchmarks considered, the FFT solver outperforms the rest in terms of absolute time. However, its strong scaling behaviour on the Alps supercomputer is limited by the poor performance of the particle update kernel, which exchanges particles with neighbouring ranks during each time step. This issue has been identified, and a fix is currently in development. 

The finite-differences based PCG solver performs competitively and is well suited for simulations requiring Dirichlet and Neumann boundary conditions, with the caveat that it is a second-order method and therefore less accurate. 

The FEM solver is advantageous in applications requiring higher accuracy, since higher-order basis functions can be used without refining the grid \cite{hiptmairNumericalMethodsPartial2024}. A matrix-free higher-order FEM solver is in development. 

For cases where accuracy is critical and we want to avoid aliasing errors present in explicit PIC, PIF is the clear choice. As shown in this work, we do not lose much in terms of performance or absolute runtime when using PIF instead of traditional PIC schemes. It can therefore present itself as a new method available to computational physicists who would like to run high-fidelity simulations with better conservation and stability properties \cite{muralikrishnan2025parapif}. However, similarly to the FFT PIC scheme, its applicability is limited to periodic and free space boundary conditions.

This study complements and aligns with the work presented in \cite{gholamiFFTFMMMultigrid2016}, where they compare different Poisson solvers, namely FFT, FMM, and multigrid, albeit not in the context of a full PIC simulation as here. Their findings show more than an order of magnitude difference between timings for the FFT solver, and the FMM and Algebraic/Geometric Multigrid solvers, reinforcing the strength of FFT under suitable conditions. The choice of method in the end should be tailored to the problem the user seeks to solve. 

Finally, matrix-free preconditioning remains a key challenge for improving the performance and scalability of the FEM solver, and will be tackled in future work.

\begin{acks}
    We acknowledge access to Alps at the Swiss National Supercomputing Centre, Switzerland under the Paul Scherrer Institute's share with the project ID psi07 and the PASC project with ID c41, as well as access to PASC consultants via this project, among which John Biddiscombe. We acknowledge CSCS, Switzerland for awarding this project access to the LUMI supercomputer, owned by the EuroHPC Joint Undertaking, hosted by CSC (Finland) and the LUMI consortium through CSCS, Switzerland. We would also like to acknowledge the LUMI porting and optimization program which organized the AMD LUMI Hackathon, that took place in December 2025, hosted by CSCS, through which we also gained performance insights and allocations on the LUMI machine. We would like to acknowledge the computing time in the JUWELS Booster supercomputer provided under the projects CSTMA and Helmholtz Association's Initiative and Networking Fund on the HAICORE@FZJ partition. Finally, we would like to thank Lukas Bühler, Jonas Meier, Matteo Bolliger and Bob Schreiner for their help in bringing the preconditioners and FEM method into IPPL.
\end{acks}

\section*{Availability}
IPPL is an open source project.\ The source code can be found here: \url{https://github.com/IPPL-framework/ippl}. The branches used for this study can be found in the following tags of IPPL: \url{https://github.com/s-mayani/ippl/releases/tag/pasc26_paper_v1} for different PIC schemes and \url{https://github.com/srikrrish/ippl/releases/tag/nufft_pasc_submission_r1} for the PIF scheme. 
On Alps, we use \texttt{gcc/13.4}, \texttt{cuda/12.8.1}, and \texttt{cray-mpich/9.0.0}.
On LUMI, we work with the LUMI software stack \texttt{LUMI/24.03} with \texttt{cray-mpich/8.1.29} and \texttt{rocm/6.0.3} for the strong scaling, and \texttt{LUMI/25.03} with \\ \texttt{cray-mpich/8.1.32} and \texttt{rocm/6.3.4} for the weak scalings (since they were done at a later date after a LUMI system update).
On JUWELS Booster, we load the following modules: \texttt{gcc/13.3}, \texttt{cuda/12}, and \texttt{OpenMPI/5.0.5}. We use Kokkos version 4.7.1 and \texttt{heFFTe} version 2.4.1.

\bibliographystyle{ACM-Reference-Format}
\bibliography{bibliography.bib,nufft.bib}

\appendix
\section{Non-uniform Fast Fourier Transform (NUFFT)} \label{appendix:nufft}
The Non-Uniform Discrete Fourier Transform (NUDFT) extends the classical DFT to situations where sample locations are not arranged on a regular Cartesian lattice. In the type~1 transform, irregularly spaced spatial samples are mapped to Fourier coefficients on a uniform mode grid. Type~2, which is the adjoint of type~1, evaluates a function represented via uniform Fourier modes at arbitrary spatial locations.

We consider the three-dimensional setting with \(N\) Fourier modes per dimension, \(N_m = N^3\) total modes, and domain length \(L\). Assuming \(N\) is even, we define $K_N = \left\{\frac{2\pi}{L}\,[-N/2, \ldots, N/2 - 1]\right\}^3$.

Given \(N_p\) non-equispaced points \(\mathbf{x}_j\), the NUDFTs of type~1 and type~2 are
\begin{align}
\label{p2f}
\hat{f}(\mathbf{k}) &= 
\sum_{j=1}^{N_p} f(\mathbf{x}_j)
\exp\!\left(-i\,\mathbf{k}\cdot\mathbf{x}_j\right), 
\qquad \mathbf{k}\in K_N ,\\[4pt]
\label{f2p} 
f(\mathbf{x}_j) &= 
\sum_{\mathbf{k}\in K_N} \hat{f}(\mathbf{k})
\exp\!\left(i\,\mathbf{k}\cdot\mathbf{x}_j\right),
\qquad j = 1,\ldots,N_p .
\end{align}
A direct computation of the NUDFT scales as \(\mathcal{O}(N_p N_m)\), which rapidly becomes impractical as the number of samples or modes increases.

Non-uniform FFT (NUFFT) algorithms reduce this cost to
\[
\mathcal{O}\!\left(|\log \varepsilon|^d\,N_p + N_m \log N_m\right),
\]
where \(\varepsilon\) is the desired accuracy and \(d\) is the spatial dimension \cite{potts2001fast,barnett2019parallel}.  
Following the formulation in \cite{pippig2016massively}, the type~1 NUFFT can be written as
\begin{equation}
\label{type1nufft}
    \hat{f} =
    D_{N_m}\,
    \chi_{N_m,N_{\tilde m}}\,
    \mathcal{F}_{N_{\tilde m},N_{\tilde m}}\,
    C_{N_{\tilde m},N_p}\,
    f .
\end{equation}
The computation of \eqref{type1nufft} proceeds as follows: first, the non-uniform data values are spread onto an oversampled Cartesian grid of size \(N_{\tilde m} = M^3\), where \(M=\sigma N\), and the oversampling factor satisfies \(\sigma>1\). In practice, \(\sigma=2\) is commonly chosen in order to suppress aliasing effects \cite{barnett2019parallel,barnett2021aliasing}. This spreading operation is performed using a compact, smooth window function, which is localized both in physical and in Fourier space, and is represented by the sparse matrix \(C_{N_{\tilde m},N_p}\). 

Subsequently, a standard uniform FFT is applied on the oversampled grid, denoted by the operator \(\mathcal{F}_{N_{\tilde m},N_{\tilde m}}\). From the resulting Fourier coefficients, the relevant \(N_m\) modes associated with the target frequency set \(K_N\) are then extracted using the characteristic selection matrix \(\chi_{N_m,N_{\tilde m}}\), which discards all remaining modes. Finally, the smoothing introduced by the window function during the spreading step is removed by applying the diagonal matrix \(D_{N_m}\), whose entries contain the Fourier transform \(\varphi_{\mathbf{k}}\) of the window function evaluated at the selected Fourier modes \(\mathbf{k}\in K_N\).

For the type~2 NUFFT, corresponding to \eqref{f2p}, these steps are reversed. The matrix form reads
\begin{equation}
\label{type2nufft}
    f = C_{N_{\tilde m},N_p}^{T}\,\mathcal{F}^{-1}\, \chi_{N_m,N_{\tilde m}}^{T}\,D_{N_m}\,\hat{f}.
\end{equation}

\end{document}